# An Item-Based Collaborative Filtering using Dimensionality Reduction Techniques on Mahout Framework

Dheeraj kumar Bokde
Department of Information Technology
Maharashtra Institute of Technology
Pune, India
bokde.dheeraj@gmail.com

Sheetal Girase
Department of Information Technology
Maharashtra Institute of Technology
Pune, India
girase.sheetal@gmail.com

Debajyoti Mukhopadhyay
Department of Information Technology
Maharashtra Institute of Technology
Pune, India
debajyoti.mukhopadhyay@gmail.com

*Abstract*—Collaborative Filtering (CF) is the most widely used prediction technique in Recommendation System (RS). Most of the current CF recommender systems maintains single criteria user rating in user-item matrix. However, recent studies indicate that recommender system depending on multi-criteria can improve prediction and accuracy levels of recommendation by considering the user preferences in multi aspects of items. This gives birth to Multi-Criteria Collaborative Filtering (MC-CF). In MC-CF users provide the rating on multiple aspects of an item in new dimensions, thereby increasing the size of rating matrix, sparsity and scalability problem. Appropriate dimensionality reduction techniques are thus needed to take care of these challenges to reduce the dimension of user-item rating matrix to improve the prediction accuracy and efficiency of CF recommender system. The process of dimensionality reduction maps the high dimensional input space into lower dimensional space.

Thus, the objective of this paper is to propose an efficient MC-CF algorithm using dimensionality reduction technique to improve the recommendation quality and prediction accuracy. Dimensionality reduction techniques such as Singular Value Decomposition (SVD) and Principal Component Analysis (PCA) are used to solve the scalability and alleviate the sparsity problems in overall rating. The proposed MC-CF approach will be implemented using Apache Mahout, which allows processing of massive dataset stored in distributed/non-distributed file system.

## I. INTRODUCTION

Nowadays the amount of information available online is increased exponentiontially information overload problem becomes challenging for Information Retrieval. In such environment customers of the e-Commerce websites have difficulty to find optimal information they need. To help customers, e-Commerce companies and researchers cameup with the solution as Recommendation System.

RS are the software tools to select the online information relevant to a given user. The first RS was developed by Goldberg, Nichols, Oki & Terry in 1992. Tapestry [16] was an electronic messaging system that allowed users to either rate messages ("good" or "bad"). Recommendation System can be classified into: Content-Based (CB), Collaborative Filtering (CF) and Hybrid Recommendation System [2][7]. RS based on CF is much explored technique in the field of Machine Learning and Information Retrieval and has been successfully employed in many applications. In CF, past user behavior are analyzed in order to establish connections between users and items to recommend an item to a user based on opinions of other users'. Customers who had similar tastes in the past, will have similar tastes in the future. CF algorithms can be further divided into two categories: memory-based algorithm (user-based and item-based) and model-based algorithm. Many e-Commerce companies have already incorporated RS with their services. Examples for such RSs include Product and Book recommendation by Amazon, Movie recommendations by Netflix, Yahoo!Movies and MovieLens, Product advertisements shown by Google based on the search history.

For large and complex data, CF methods frequently give better performance and accuracy than other RS techniques. Early CF algorithms utilizes the association inference, which have very high time complexity and very poor scalability. While recent methods that use matrix operations are more scalable and efficient [1-2]. The implementation CF algorithms in the applications of RSs face several challenges. First is the size of processed datasets. The second one comes from the sparseness of referenced data, which means missing value in user-item rating matrix. Also, memory-based CF approaches suffers from the scalability problem. Compared to memory-based algorithms, model-based algorithms scales better in terms of resource requirements. Several research has suggested that model-based CF provide better predictive accuracy than memory-based CF, by using techniques such as Matrix Factorization (MF) and Dimentionality Reduction [2][8].

In this paper we proposed a new approach to improve the predictive accuracy and efficiency of Multi-Criteria Collaborative Filtering using Dimensionality Reduction techniques and its Mahout [3-5][15] implementation for a recommendation system application. It should be noted that although our algorithm is designed for item-based CF approach [6] considering multi-criteria features, it can be modified to become a user-based method. In proposed approach PCA option for Higher Order SVD is used for scalability improvement, preciseness improvement and predicting the missing values in user-item rating matrix.

### A. Organization

The rest of this paper is organized as follows. In Section II, MC-CF is presented. In Section III, Apache Mahout is introduced. In Section IV, MC-CF based on

SVD, SSVD, higher order SVD and PCA are discussed. In Section V, proposed research methodology is presented. In Section VI, experimental dataset and metrics to evaluate the predective accuarcy and performance of our implementation are described. The conclusion is given in the last section.

## II. MULTI-CRITERIA ITEM-BASED COLLABORATIVE FILTERING

Before introducing MC-CF, a brief information of traditional CF algorithm is needed. Here we explain the item-based CF algorithm proposed by Sarwar et al. [6]. In a single rating CF system the collected user rating is utilized to predict a rating given by a function:

$$R: Users \times Items \rightarrow R_0$$

Where, $R_0$ is the set of possible overall ratings

The task of prediction was accomplished in the item-based CF by forming each item's neighborhood, Sarwar et al. proposed the adjusted cosine-based similarity as a measure for estimating items distance. After similarities are computed then first forms the nearest neighborhood of an item $i$, then calculate prediction for an active user for a target item [6-7].

Traditional single criteria rating-based CF system requests an overall rating from users in oreder to collect their preferences. Then in multi-criteria rating-based CF system there is more than one criterias, therefore traditional prediction modified as follows:

$$R: Users \times Items \rightarrow R_0 \times R_1 \times R_2 \times ..... \times R_k$$

Where, $R_0$ is the set of possible overall ratings
$R_i$ indicates the possible rating for each criteria $i$
$k$ shows the number of criteria

In [13] mentioned, according to Adomavicius and Kwon, MC-CF recommender systems predict the overall rating for an item based on past ratings regarding both the item overall and individual criteria, and recommend to the user the item with the best overall score. Thus, the algorithm for a MC-CF recommender system can be extended from a single-rating recommender system. In MC-CF problem, there are $m$ users, $n$ items and $k$-criteria in addition to the overall rating. Users have provided a number of explicit ratings for items, a general rating $R_0$ must be predicted in addition to $k$ additional criteria ratings $(R_1, R_2 . . . , R_k)$.

The MC-CF process can be defined in two steps, by predicting the target user's or rating for a particular unseen item, followed by producing a Top-N list of recommendations for a given target. Recommendation is a list of Top-N items, $T = \{T_1, T_2, . . . , T_N\}$, that the active user will like the most [12-13]. The recommended list usually consists of the products not already purchased by the active customer.

## III. APACHE MAHOUT

Apache Mahout is an open source Machine Learning library which provides an efficient framework utility for distributed/non-distributed programming [15]. It is scalable and can handle large amount of data compared to other Machine Learning framework. Mahout was started in 2008 as a spin off technology from the Apache Lucene project, which was primarily concerned with content search and Information Retrieval technologies. Apache Mahout is one of the framework in Apache Hadoop [16] projects. Mahout contains three types of algorithms: Recommender System (specially Collaborative Filtering), Clustering and Classification. Mahout has its own seprate open source project called "Taste" for Collaborative Filtering. The implementations of recommender systems can be further categorized as non-distributed methods and distributed methods. The distributed implementation of RS uses MapReduce, which is scalable and suitable to handle massive and distributed dataset. Its scalability and focus on real world applications makes Mahout an increasingly popular choice for organizations seeking to take advantage of large scale Machine Learning techniques.

The Apache Mahout architecture for non-distributed recommender engine is shown in the Fig. 1, provides a rich set of components from which we can construct a customized Recommender System by the selection of algorithms. Mahout is designed to be enterprise-ready designed for performance, scalability and flexibility. Top-level packages define the Apache Mahout interfaces to these key abstractions [3][5][15] are:

- DataModel
- UserSimilarity
- ItemSimilarity
- UserNeighborhood
- Recommender

Mahout provides these building blocks from which we can construct the best Recommender for our application. Fortunately, Mahout also provides a way to evaluate the accuracy of our Recommenders on datasets using header org.apache.mahout.cf.taste.eval.

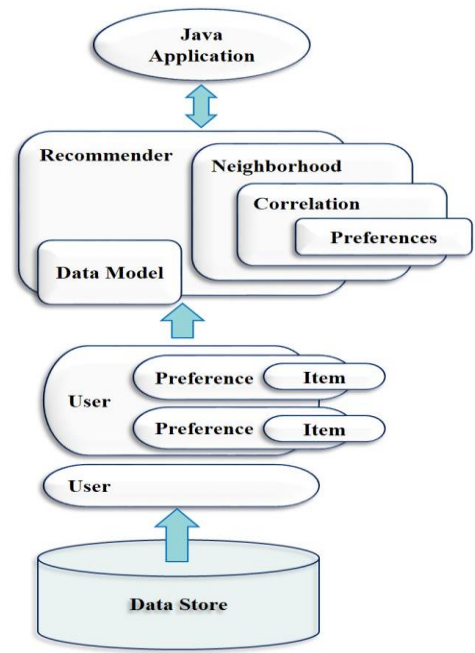

Figure 1. Apache Mahout Architecture [15].

## IV. MULTI-CRITERIA COLLABORATIVE FILTERING USING DIMENTIONALITY REDUCTION TECHNIQUES

To address the challenges of MC-CF algorithms like scalability and sparsity, many researchers have used the Dimensionality Reduction Techniques with CF algorithms. The Memory-Based CF algorithms usually uses similarity metrics to obtain the similarity between two users, or two items based on each of their ratios. CF algorithms can be further divided into user-based and item-based approaches. This section explores the user-based CF and item-based CF as well as their implementation with Dimensionality Reduction [8-11].

In general, the process of dimensionality reduction can be described as mapping a high dimensional input space into a lower dimensional latent space. A special case of dimensionality reduction is matrix factorization where a data matrix D is reduced to the product of several low rank matrices. Dimensionality Reduction techniques are:

- Singular Value Decomposition
- Principal Components Analysis

### A. Singular Value Decomposition

A powerful technique for dimensionality reduction is SVD and it is a particular realization of the MF approach. To find a lower dimensional feature space is the key issue in a SVD. SVD of matrix $A$ of size $m \times n$ is of the form [9-10]:

$$SVD(A) = U \Sigma V^T \quad (1)$$

Where, $U$ and $V$ are $m \times m$ and $n \times n$ orthogonal matrices respectively, $\Sigma$ is the $m \times n$ singular orthogonal matrix with non-negative elements.

An $m \times m$ matrix $U$ is called orthogonal if $U^T U$ equals to an $m \times m$ identity matrix. The diagonal elements in $\Sigma(\sigma 1, \sigma 2, ...., \sigma n)$ are called the singular values of matrix $A$. Usually, the singular values are placed in the descending order in $\Sigma$. The left and right singular vectors are the column vectors of $U$ and $V$ respectively. Many important applications use SVD's desirable properties. Amongst them is the low rank approximation of matrix. The truncated SVD of rank $k$ is defined as [9-10]:

$$SVD(A_k) = U_k \Sigma_k V_k^T \quad (2)$$

Where, $U_k$ and $V_k$ are $m \times k$ and $n \times k$ matrices composed by the first $k$ columns of matrix $U$ and the first $k$ columns of matrix $V$ respectively, Matrix $\Sigma_k$ is the $k \times k$ principle diagonal sub-matrix of $\Sigma$. $A_k$ represents the closest linear approximation of the original matrix A with reduced rank $k$. Once the transformation is completed, user and items can be thought off as points in the $k$-dimensional space.

### B. Stochastic Singular Value Decomposition

The high computational cost is the biggest problem of using SVD (described above) in CF algorithms. When the target matrix A is large and sparse and only the low rank approximation is desired the Krylov subspace method can be used [15]. The hardness of parallelizing the Krylov subspace methods and the variant number of iterations become the critical deficiency of processing the massive datasets. The SSVD algorithm is shown in Fig. 2 below

---

**Input:** An $m \times n$ matrix $A$, and a number $k$
**Output:** Approximate $U_k$, $V_k$ and $\Sigma_k$
**Algorithm:**
1. Generate an $n \times k$ Guassian matrix $G$
2. Compute $Y = AG$
3. Compute an orthogonal column basis $Q$ of $Y$
4. Form $B = Q^T A$
5. Compute eigen-decomposition of $BB^T = X\Sigma^2 X^T$
6. $U_k = QX$, $V_k = B^T X \Sigma^{-1}$ and $\Sigma_k = \Sigma$

Figure 2. Stochastic Singular Value Decomposition Algorithm [1].

### C. Higher Order Singular Value Decomposition

The limitation of using SSVD Item-Based CF algorithm is that, it is potentially less precise and can be applied only to two dimensional user-item rating matrix. When we consider multiple criterias for rating purpose, the user-item matrix becomes three dimensional to handle this problem we need higher order SVD. Higher Order Singular Value Decomposition (HOSVD) or Multilinear SVD was proposed by Lathauwer at el. (2000) [17], is a generalization of SVD that can be applied on three (or more) dimension called Tensor. The objective is to compute low-rank approximation of the data. These approximations are exressed in terms of tensor decomposition. The HOSVD of a third-order tensor involves the computation of the SVD of three matrices called modes.

The SVD of a real matrix A is give by: $A = U\Sigma V^T$. In HOSVD, S is in generally not sparse or diagonal as $\Sigma$ in the SVD. The HOSVD of third order tensor of $A \in R^{I_1 \times I_2 \times I_3}$ for every $I_1 \times I_2 \times I_3$, can be written as the product:

$$A = S \times_1 U^{(1)} \times_2 U^{(2)} \times_3 U^{(3)} \quad (3)$$

The core tensor S can then be computed (if needed) by bringing the matrices of s-mode singular factors to the left side of equation (4):

$$S = A \times_1 U^{(1)H} \times_2 U^{(2)H} \times_3 U^{(3)H} \quad (4)$$

Where $(.)^H$ denotes the complex conjugate.

### D. Principal Components Analysis

The powerful technique of dimensionality reduction is PCA and it is a particular realization of the MF approach [2][11]. PCA is a statistical procedure which uses an orthogonal transformation that converts a set of observations of possibly correlated variables into a set of values which are linearly uncorrelated variables called Principal Component (PC). The number of PC is less than or equal to the number of original variables. This transformation is defined in such a way that the first PC has the largest possible variance and each succeeding component in turn has the highest variance possible under the constraint that it is orthogonal to the preceding components. The PC are orthogonal because they are the eigenvectors of the covariance matrix. PCA is sensitive to the relative scaling of the original variables. Algorithmic steps for PCA using eigenvalue and eigenvectors is shown in the Fig. 3.

**Algorithm:**
Step 1: Get the data from $m \times n$ matrix $A$
Step 2: Calculate the covariance matrix
Step 3: Calculate the eigenvectors and eigenvalues of the covariance matrix
Step 4: Choosing principal components and forming a feature vector
Step 5: Deriving the new data set and forming the clusters

Figure 3. Algorithmic steps of Principal Components Analysis [11].

## V. PROPOSED METHODOLOGY

The Item-based CF algorithm using SSVD not only provide accurate results but also reduces the computational cost. The potential limitation of SSVD, that it is potentially less precise and it can't be applied directly to three dimensional user-item rating matrix with multiple criteria. So, to overcome this challenge we propose an idea to use PCA option for Higher Order SVD based Collaborative Filtering algorithm. PCA (described in section IV.D) finds a linear projection of high dimensional data into a lower dimensional subspace such as the variance retained is maximized and the least square reconstruction error is minimized. Fig. 4 shows the proposed architecture of CF approach.

Recall that the PCA technique uses the rank-k of SVD method to order the input data in descending order of importance and correlation. This way the most important and less uncorrelated input components are given higher priority than the less important and highly correlated ones. In the context of massive computation input rating matrix is often sparse. Thus, sparser the original input, the more efficiency gain by using PCA option for Higher Order SVD as compared to SSVD approach. This able to find the potentially precise recommendations for the user. PCA is also used in the computation of missing valuses in the user-item rating matrix.

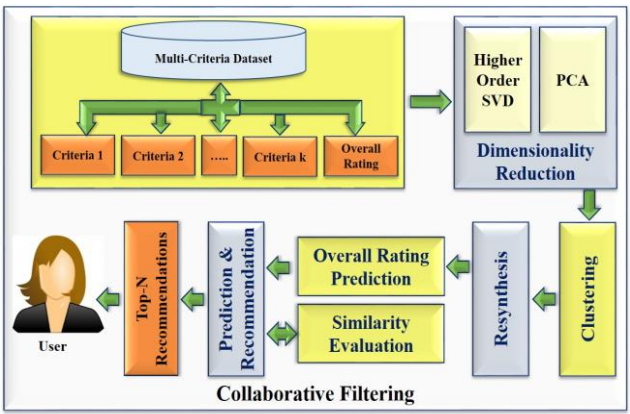

Figure 4. Proposed Collaborative Filtering Architecture.

The components of proposed Collaborative Filtering approach are:
- Multi-Criteria Dataset
- Dimensionality Reduction and Clustering
- Resynthsis and Overall Rating Prediction
- Similarity Evaluation
- Prediction and Recommendation

### A. Multi-Criteria Dataset

CF recommender systems predict the overall rating for an item based on past ratings regarding both the item overall and individual criteria, and recommend an item to the user having best overall score [12-13]. Nowadays e-Commerce industry is growing exponentiontially, thus information overload becomes a challenge for Information Retrival. In such environment users of the e-Commerce sevices face difficulty to find optimal information what they need. As the users are using these systems their changing needs, selection of items based on multiple criteria's and variety of products is making business more and more complex. To better understand the users prefernce for recommendation, there is need to include multi-criteria fatures with the RS.

### B. Dimentionality Reduction and Clustering

In general, the process of dimensionality reduction can be described as mapping a high dimensional input space into a lower dimensional latent space. A special case of dimensionality reduction is matrix factorization where a data matrix D is reduced to the product of several low rank matrices. Dimensionality Reduction techniques like SVD, PCA and tensor decomposition are discribed in the Section III. As a result of dimentionality reduction step we get the clusters of the correlated datasets. Because in the PCA will classify the rating matrix with new dimension.

### C. Resynthesis and Overall Rating Prediction

After the clustering we have to compute the overall rating of the active items for the active users. We present a method for analyzing and re-synthesizing in homogeneously textured regions in matrix. PCA has been used for analytic purposes but rarely for matrix compression. This is due to the fact that the transform has to be defined explicitly. The correlated singular item matrix obtained may contain some noisy data which is of no importance and may increase the computation cost of the algorithm. So we need to apply an efficient resynthesis method to filter those noisy data from the singular item matrix. For this purpose we use the Fuzzy Rules to compute rating prediction.

### D. Similarity Evaluation

Now we have to perform similarity evaluation using the multiple-criteria features over new dimension matrix. The similarity evaluation can be done by finding the distance between the item vectors using distance measures like: Manhatten distance, Euclidean distance and Chebyshev distance etc.[13], and the resultant distance can be converted to the similarity value. There are various similarity metrics like Cosine similarity, Pearson similarity etc. We can use these similarity metrics to find the similarity between the item vectors considering the multi-criteria. Finally we have to build the correlation between those items to compute prediction.

## E. Prediction and Recommendation

After filtering the noisy data from the item matrix by Resynthesis method and similarity evaluation now we have to perform prediction operation. The CF recommender system follows the prediction and recommendation task of MC-CF recommender system using three main steps as [12-13]:

Step 1: Selecting the active user and active items

Step 2: Predicting criteria ratings using the neighborhood formation on individuals, which solves the sparsity problem

Step 3: Predicting the overall rating for the active users and recommending the Top-N items to user according to users' preference considering multi-criteria

## VI. EXPERIMENTS AND EVALUATION METRICS

### A. Experimental Dataset

In order to analyse the effectiveness of the proposed method we are going to perform experiments using Yahoo!Movies dataset provided by Yahoo!Research Alliance Webscope program available on the website: (http://webscope.sandbox.yahoo.com).

On the Yahoo!Movies network users could rate in 4 dimensions (Story, Acting, Direction and Visuals) and assign an overall rating. These four features of any movies we consider as multi-criteria. Users use a 13-level rating scale (from A+ to F) for rating purpose. In the original dataset there are 257,317 rating, with 127,829 users and 8272 movies. For processing purpose, we replaced letters with numbers so as 1 correspond to worst value denoted as F in original dataset. After pre-processed the dataset we created the test datasets with different density and quality level into YM-20, YM-10, YM-5 on the basis that users and movies have atleast 20, 10, 5 ratings respectively, is presented in the Table I.

TABLE I. INFORMATION OF YAHOO!MOVIES DATASET

| Name | # Users | # Items | # Overall Ratings |
|---|---|---|---|
| YM-20 | 429 | 491 | 18,405 |
| YM-10 | 1,827 | 1,471 | 48,026 |
| YM-5 | 5,978 | 3,079 | 82,599 |

### B. Evaluation of Recommendation System

After almost two decades of research on CF algorithms many researchers came up with various evaluation metrics. To measure the quality of proposed method this section presents the various evaluation metrics used to evaluate the prediction accuracy, performance and effective implementation of the proposed algorithm and its Mahout implementation.

#### 1) Coverage

Coverage is defined as the percentage of items the Recommendation Systems able to recommend to the user. Coverage can be used to detect algorithms accuracy, although they recommend only a small number of items. These are usually very popular items with which the user is already familiar without the help of the system [7].

#### 2) Prediction Accuracy

Prediction accuracy metrics measures the recommender's predictions that are close to the true users rating. There are various metrics used by the CF researchers to check the prediction accuracy of their implemented algorithms are Mean Absolute Error (MAE), Root Mean Squared Error (RMSE), Precision, Recall and F1 Metric. Formulas of these metrics are [1][7][13]:

$$MAE = \frac{\sum_{i}^{k}(p_i - r_i)}{k} \quad (5)$$

$$RMSE = \sqrt{\frac{\sum_{i}^{k}(p_i - r_i)^2}{k}} \quad (6)$$

$$Precision = \frac{|Interesting\ Items \cap Recommended\ Items|}{Recommended\ Items} \quad (7)$$

$$Recall = \frac{|Interesting\ Items \cap Recommended\ Items|}{Interesting\ Items} \quad (8)$$

$$F1\ Metric = \frac{2(Recall \times Precision)}{Recall + Precision} \quad (9)$$

Where,
- $p_i$ is the prediction for user $i$
- $r_i$ is the prediction for how user $i$ will rate item i.e., real or true rating value
- $k$ is the number of items user $i$ has rated i.e., {user, item} pair

With the help of these evaluation metrics we can calculate and compare the prediction accuracy and efficiency of proposed CF algorithm with existing approaches and decide which algorithm performs better.

### C. Experiments

The first experiment is concerned with evaluating Item-Based CF using similarity algorithms implemention using the Apache Mahout with the GroupLens' MovieLens 100k dataset available on (http://grouplens.org/datasets/movielens/). Detailed information of MovieLens 100k Dataset is shown in Table II. This collection of dataset includes data about movies including users, movies, and movie ratings from users.

TABLE II. INFORMATION OF MOVIELENS 100K DATASET

| Name | # Users | # Items | #Preferences |
|---|---|---|---|
| ML 100k | 943 | 1,682 | 100,000 |

TABLE III. EVALUATION OF RMSE AND MAE FOR ITEM-BASED CF USING SIMILARITY MEASURES ON ML 100K DATASET

| Item-Based Collaborative Filtering | | | | |
|---|---|---|---|---|
| Similarity Measure | MAE | | RMSE | |
| | 70% Training | 80% Training | 70% Training | 80% Training |
| Pearson Correlation Similarity | 0.842 | 0.828 | 1.080 | 1.061 |
| Euclidean Distance Similarity | 0.816 | 0.818 | 1.022 | 1.026 |
| Log Likelihood Similarity | 0.814 | 0.817 | 1.019 | 1.025 |
| Tanimoto Coefficient Similarity | 0.793 | 0.794 | 0.999 | 1.002 |

The experiments are evaluated for Item-Based CF that utilize the Pearson Correlation Coefficient, Log Likelihood similarity, Euclidean Distance Similarity and Tanimoto Coefficient Similarity algorithms. The purpose of this experiment is to determine how similarity measures affect the prediction accuracy of an Item-Based CF algorithm. The following results are obtained using MovieLens 100k, when used 70 and 80 percent training data over ML 100k dataset. Tables III represent evaluation of Item-Based CF algorithms with different similarity algorithms to evaluate prediction accuracy.

## VII. Conclusion

In this paper, a method of Collaborative Filtering with Dimensionality Reduction technique using Mahout is proposed to improve the recommendation quality and predictive accuracy. We have proposed this method for overcoming the existing shortcomings such as predicting the overall rating, sparsity, scalability, imprecision and massive dataset features in Multi-Criteria CF.

Using dimensionality reduction techniqus such as SVD and PCA we reduce the noise of high dimentional data, improve the scalbilty and tackle the sparsity problem of rating matrix. The SSVD not only reduces the computation cost of Multi-criteria Item-Based CF algorithm but also increases the accuracy and efficiency of the MC-CF algorithms. The potential limitation of SSVD is that it is potentially less precise and can't be directly applied to three dimensional rating matrix. So, to overcome this challenge we propose an idea to use PCA option for Higher Order SVD based CF algorithm with multi-criteria features using Mahout.

## References

[1] Che-Rung Lee, Ya-Fang Chang, "Enhancing Accuracy and Performance of Collaborative Filtering Algorithm by Stochastic SVD and Its MapReduce Implementation," *Published in IEEE 27th International Symposium on Parallel & Distributed Processing Workshops and PhD Forum*, IEEE Computer Society, IEEE 978-0-7695-4979-8/13, ©IEEE, 2013.

[2] Francesco Ricci, Lior Rokach, Bracha Shapira and Paul B. Kantor, "Recommender Systems Handbook," *Springer*, ISBN: 978-0-387-85819-7, ©Springer Science + Business Media LLC, 2011.

[3] Sebastian Schelter, Sean Owen, "Collaborative Filtering with Apache Mahout," ACM RecSysChallenge'12, Dublin, Ireland, September 13, 2012.

[4] Carlos E. Seminario, David C. Wilson, "Case Study Evaluation of Mahout as a Recommender Platform," ACM RecSys Challenge'12, Dublin, Ireland, pp. 45-50, September 13, 2012.

[5] Sachin Walunj, Kishor Sadafale, "An Online Recommendation System for E-commerce Based on Apache Mahout Framework," ACM SIGMIS-CPR'13, 978-1-4503-1975-1/13/05, Cincinnati, Ohio, USA, May-30-June-1, ©ACM, 2013.

[6] Sarwar B., Karypis G., Konstan J., Riedl J., "Item-based Collaborative Filtering Recommendation Algorithms," *Published in the Proceedings of the 10th international conference on World Wide Web*, Hong Kong, ACM 1581133480/01/0005, ©ACM, May 15, 2001.

[7] Fidel Cacheda, Victor Carneiro, Diego Fernandez, and Vreixo Formoso, "Comparison of Collaborative Filtering Algorithms: Limitations of Current Techniques and Proposals for Scalable, High-Performance Recommender Systems," *ACM Transactions on the Web*, Vol. 5, No. 1, Article 2, ©ACM, February 2011.

[8] Yehuda Koren, "Matrix Factorization Techniques for Recommender Systems," *Published by the IEEE Computer Society*, IEEE 0018-9162/09, pp. 42- 49, ©IEEE, August 2009.

[9] M.G. Vozalis and K.G. Margaritis, "Using SVD and demographic data for the enhancement of generalized Collaborative Filtering," *Published in An International Journal of Information Sciences 177(2007)*, 0020-0255, pp. 3017-3037, © Elsevier Inc., February 2007.

[10] SongJie Gong, HongWu Ye and YaE Dai, "Combining Singular Value Decomposition and Item-based Recommender in Collaborative Filtering," *Second International Workshop on Knowledge Discovery and Data Mining,* 978-0-7695-3543-2/09, pp. 769-772, © IEEE, 2009.

[11] Manolis G. Vozalis and Konstantinos G. Margaritis, "A Recommender System using Principal Component Analysis," *Published in 11th Panhellenic Conference in Informatics*, Patras, Greece, pp. 271-283, 18-20 May, 2007.

[12] Wiranto, Edi Winarko, Sri Hartati and Retantyo Wardoyo, "Improving the Prediction Accuracy of Multicriteria Collaborative Filtering by Combination Algorithms," *Pubilshed in (IJACSA) International Journal of Advanced Computer Science and Applications, Vol. 5, No. 4, pp. ,52-58, 2014*.

[13] Alper Bilge and Cihan Kaleli, "A Multi-Criteria Item-Based Collaborative Filtering Framework," *Published in 11th (JCSSE) International Joint Conference on Computer Science and Software Engineering,"* IEEE 978-1-4799-5822-1/14, © IEEE, 2014.

[14] David Goldberg, David Nichols, Brian M. Oki and Douglas Terry, "Using Collaborative Filtering to Weave an Information Tapestry," *Communications of the ACM*, Vol. 35, No.12, December 1992.

[15] Apache Mahout, Available at URL: http://mahout.apache.org/

[16] Apache Hadoop, Available at URL: http://hadoop.apache.org/

[17] R Boyer, R Badeau, "Adaptive Multilinear SVD for Structured Tensors," Published in the *Proceedings IEEE International Conference on Acoustics, Speech and Signal Processing ICASSP 2006*, vol. no. 3, pp .III, 14-19 May 2006.